\documentclass[
 reprint,
 superscriptaddress,
 amsmath,
 amssymb,
 aps,
 pra,
]{revtex4-2}

\usepackage[english]{babel} 
\usepackage[utf8]{inputenc}
\usepackage[T1]{fontenc}
\usepackage{hyperref}
\usepackage{url}
\usepackage{booktabs}
\usepackage{amsfonts}
\usepackage{nicefrac}
\usepackage{microtype}
\usepackage{graphicx}
\usepackage{svg}
\usepackage{amsmath}
\usepackage{mathrsfs}

\begin{document}


\title{Spontaneous optical skyrmion generation by frequency doubling in underdense plasmas}

\author{Marcos G. Barriopedro} \email{m.gbarriopedro@upm.es}
\affiliation{Instituto de Fusión Nuclear "Guillermo Velarde", ETSII, Universidad Politécnica de Madrid, José Gutiérrez Abascal 2, 28006 Madrid, Spain}
\affiliation{Complex Systems Group, ETSIME, Universidad Politécnica de Madrid, Ríos Rosas 21, 28003 Madrid, Spain}

\author{Eduardo Oliva}
\affiliation{Instituto de Fusión Nuclear "Guillermo Velarde", ETSII, Universidad Politécnica de Madrid, José Gutiérrez Abascal 2, 28006 Madrid, Spain}
\affiliation{Departamento de Ingeniería Energética, ETSII, Universidad Politécnica de Madrid, José Gutiérrez Abascal 2, 28006 Madrid, Spain}

\author{Miguel A. Porras}
\affiliation{Complex Systems Group, ETSIME, Universidad Politécnica de Madrid, Ríos Rosas 21, 28003 Madrid, Spain}

\begin{abstract}
    Optical skyrmions have been widely explored in recent years. Among them, Stokes skyrmions require sophisticated wave engineering or photonic devices for their generation. We show that Stokes skyrmions can emerge spontaneously in second-order harmonic generation in underdense plasmas driven by optical vortices. The nonlinear response produces a structured frequency-doubled field whose polarization texture maps the Poincaré sphere. When plasma inhomogeneities are taken into account, the electron density gradient deforms the skyrmionic texture, enabling topological diagnosis of plasma density. 
\end{abstract}


\maketitle

A basic component of various forms of structured light are optical vortices (OVs) \cite{allen_orbital_1992, shen_optical_2019, berry_singularities_2023, coullet_optical_1989}. They possess a phase that varies linearly in azimuth in cross sections to the propagation direction. The number of times it varies by $2\pi$ in one turn is named topological charge (TC)  \cite{lu_topological_2014}. The vortex core, where the intensity vanishes, constitutes a phase singularity and is typically embedded in light beams such as Laguerre-Gauss (LG) or Bessel beams. They have become indispensable for a wide range of applications such as optical manipulation of matter \cite{tao_fractional_2005, cojoc_laser_2005}, information storage and transmission \cite{celechovsky_optical_2007}, quantum entanglement \cite{dambrosio_entangled_2016}, and more recently, in plasma diagnosis \cite{lopez_diagnosing_2025, lopez_conservation_2023}.

More complex forms of structured light, from cylindrical vector beams \cite{zhan_cylindrical_2009} to other generic polarization textures \cite{bauer_observation_2015}, emerge when the vector nature of light is considered. Optical skyrmions \cite{ye_theory_2024, shen_optical_2024, sugic_particle-like_2021} constitute a paradigmatic example in which a 3D vector field, parametrized on a the surface of a sphere, is mapped onto a 2D plane ($S^2\rightarrow\mathbb{R}^2$). The vector field can represent various physical observables, such as the electromagnetic field components \cite{zeng_tightly_2024, tsesses_optical_2018}, the angular spin momentum \cite{lei_photonic_2021, du_deep-subwavelength_2019}, or the Poynting vector \cite{wang_topological_2024}. The number of times that the parametric sphere is mapped on the plane is known as the skyrmionic charge or number. Among optical skyrmions, Stokes skyrmions, made with the Stokes vector on the Poincaré sphere (PS), can propagate, e.g., the full-Poincaré beam \cite{beckley_full_2010}. They are typically constructed by superposing propagation modes, e.g., LG beams of different TC and orthogonal polarizations, either with standard bulk optics or photonic devices \cite{shen_generation_2022, lin_microcavity-based_2021}. An exception is the recently reported skyrmion naturally present around the core of OVs \cite{mata-cervera_skyrmionic_2025}.

An open question is whether skyrmions can emerge from light-matter interactions \cite{rivera_lightmatter_2020} without relying on wave engineering. Recently, Marco et al. \cite{marco_attosecond_2025} have demonstrated that high-order harmonic generation driven by structured vector beams can produce attosecond extreme-ultraviolet skyrmions. In a related context, nonlinear plasma dynamics provides an alternative pathway for the emergence of topological field structures. While harmonic generation in gaseous media or nonlinear crystals depends on the individual local response of each atom, in an underdense plasma, the fluid-like electron motion results in an inherently nonlocal and directional response. It is precisely this collective motion that could allow the structure of the driving field to be transformed into a more complex one. When exposed to a spatially inhomogeneous light field, electrons migrate toward regions of lower intensity, imparting an intrinsically directional character to emitted harmonic signals, such as a second-order harmonic (SH) field. Furthermore, inhomogeneous plasma density profiles play a role in the emission, as variations in electron density can reshape the resulting SH field.

In this Letter we show that Stokes skyrmions can spontaneously emerge as the SH field generated by the nonlinear response of free electrons in an underdense plasma driven by optical vortices. When plasma can be treated as homogeneous, the second-order response produces a transversaly confined skyrmionic polarization texture with a skyrmionic charge of two, independently of the specific beam profile or the TC of the driving vortex, which evolves into a second-order meron in the far field \cite{shen_optical_2024}. When plasma inhomogeneity effects are considered, the skyrmionic polarization textures subsist in their key properties such as the skyrmionic charges, but become substantially deformed. This deformation strongly depends in the inhomogeneous plasma density profile. Thus, plasma inhomogeneities are imprinted onto the emitted field, which opens a new route for a topological diagnosis of the plasma density.

Stokes skyrmions are typically constructed by superposing two propagation modes with different TCs ($l_1$, $l_2$) and orthogonal polarizations, say $\mathbf{E} = |\mathrm{LG}_{\ell_1,0}\rangle \mathbf{u}_L + |\mathrm{LG}_{\ell_2,0}\rangle \mathbf{u}_R$, where $\mathrm{LG}_{\ell,p}$ denotes the LG beam of azimuthal order $\ell$ and radial order $p$, and $\mathbf{u}_{L,R}$ the unit vector of right- and left-handed circular polarization (RCP, LCP), respectively. The skyrmionic charge of the Stokes parameters $\mathbf{n} = (S_1,S_2,S_3)/S_0$ in the normalized PS is calculated as \cite{ye_theory_2024, shen_optical_2024}
\begin{equation} \label{eq:nsk}
    N_{sk} = \frac{1}{4\pi}
    \int_0^{r_{\max}}\!\!\int_0^{2\pi}
    \mathbf{n}\cdot
    \left(\frac{\partial\mathbf{n}}{\partial r}
    \times
    \frac{\partial\mathbf{n}}{\partial\phi}\right)
    \,dr\,d\phi,
\end{equation}
and yields $N_{sk} = \ell_2-\ell_1$, when integrating across the entire transverse plane, $r_{max}\rightarrow\infty$. In this context, generation of SH field in underdense plasmas driven by an optical vortex provides a physical mechanism for such emergence since the generated SH field naturally decomposes into a superposition of two vortex with different TCs and opposite circular polarizations.

When a sufficiently intense laser pulse propagates through a gas, the medium is ionized, forming a plasma channel. The freed electrons collectively interact with the pump pulse, leading to the generation of harmonic signals. The second-order nonlinear polarization response due to the free electrons is given by \cite{shen_principles_1984, beresna_high_2009}
\begin{equation}
    \mathbf{P}_{2\omega} = \chi(2\omega)\left[\frac{1}{2}\nabla \mathbf{E}_0^2 + \frac{2}{\varepsilon_p}\left(\mathbf{E}_0 \cdot \nabla \ln n_e\right)\mathbf{E}_0\right],
    \label{eq: P2}
\end{equation}
where $\chi(2\omega)=n_e e^3/(4 m_e^2 \omega^4)$, $e$ and $m_e$ are the electron charge and mass, $n_e$ is the electron density, $\varepsilon_p = 1 - \omega_p^2/\omega^2$ is the plasma dielectric function, with $\omega_p = \sqrt{4\pi n_e e^2/m_e}$ the plasma frequency, and $\mathbf{E}_0$ is the complex envelope of the pump pulse electric field. This second-order polarization consists of a first term related to the ponderomotive force, driven by electron oscillation toward regions of lower pump, and a second term that depends on electron density gradients.
 
To obtain the components of the SH electric field at the near field, we assume phase-matching and an underdense and weakly dispersive plasma at the driving frequency, so that propagation effects and pump depletion can be neglected over the interaction length. Within this approximation, the generated SH at the near field scales linearly with the nonlinear  polarization as \cite{boyd_chapter_2008}
\begin{equation}\label{eq:SH}
    \frac{d \mathbf{E}_{2\omega}}{\partial z} = 2i\frac{\mu_0\omega^2}{k_{2\omega}}\mathbf{P}_{2\omega}
\end{equation}
where $k_{2\omega}$ is the wavenumber of the second harmonic, and $z$ is the propagation distance.

The driving field is assumed to be a linearly polarized optical vortex of TC $\ell$,
\begin{equation}
    \mathbf{E}_0 = E_0 \left( \frac{r}{\sigma} \right)^{|\ell|} f\left( \frac{r}{\sigma} \right) e^{i\ell\phi}\, \mathbf{u}_x,
    \label{eq: E_0}
\end{equation}
where $(r,\phi)$ are polar coordinates, $\mathbf{u}_x$ is the unit vector of the direction $x$, $f(r/\sigma)$ defines the radial profile hosting the vortex, and $E_0$ and $\sigma$ set the intensity and beam width. 

\begin{figure}
\centering
\includegraphics[width=1\linewidth]{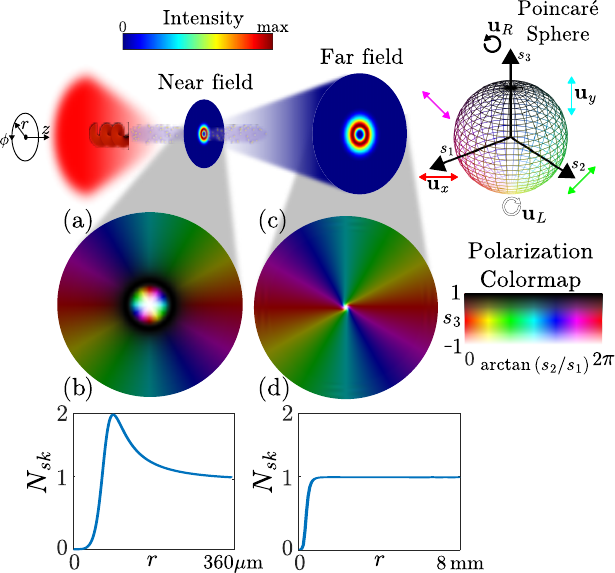}
    \caption{Upper panel: (left) A linearly polarized vortex pump pulse generates an underdense plasma and produces a Stokes skyrmion at doubled frequency covering twice the PS (right). The pump is a LG beam ${\mathbf E}_0 \propto
    \left(r/\sigma\right)\exp\left(-r^2/2\sigma^2+i\phi\right)\mathbf{u}_x$ of TC $\ell=1$, with $\sigma =80\,\mu$m and $\lambda = 800$ nm. The latitude on the PS is in gray scale and the meridian in color scale. (a) Near field mapping of the Stokes parameters onto the transversal plane. (b) Skyrmion charge in a disc of radius $r$. (c) and (d) represent the same variables after $z=10\,z_r$ cm propagation.}
    \label{fig: FIG1}
\end{figure}

To isolate the fundamental mechanism of skyrmion generation, we first neglect plasma inhomogeneities ($\nabla n_e = 0$). By neglecting the second term in   (\ref{eq: P2}), the nonlinear response is governed solely by the gradient of the driving field. Substituting equation (\ref{eq: E_0}) in equation (\ref{eq: P2}), the generated SH field $\mathbf{E}_{2\omega}$, computed from equation (\ref{eq:SH}), is readily seen to decompose into radial and azimuthal components:
\begin{equation}\label{eq:SH2}
    \mathbf{E}_{2\omega} \propto \left(\frac{r}{\sigma}\right)^{2|\ell|}f^2e^{i2\ell\phi}\left[\left(\frac{1}{f}\frac{d f}{d r}+\frac{|\ell|}{r}\right)\mathbf{u}_r + \frac{i\ell}{r}\mathbf{u}_\phi\right],
\end{equation}
where $\mathbf{u}_r$ and $\mathbf{u}_\phi$ are unit radial and azimuthal vectors, and unessential constants are omitted for brevity. For a $\ell=0$ pump the emission is purely radially polarized. However, with a vortex driver ($\ell \neq 0$), the azimuthal component strongly deforms the radial polarization, producing a singular core that encompasses the full set of polarization states on the PS. To characterize this polarization texture, we express $\mathbf{E}_{2\omega}$ in equation (\ref{eq:SH2}) in the circular polarization basis as
\begin{multline}
    \mathbf{E}_{2\omega} \propto \left(\frac{r}{\sigma}\right)^{2|\ell|}f^2e^{2i\ell\phi}\left[\left(\frac{1}{f}\frac{d f}{d r}+\frac{|\ell| - \ell}{r}\right)e^{i\phi}\mathbf{u}_R \right. \\
    \left. + \left(\frac{1}{f}\frac{d f}{d r}+\frac{|\ell| + \ell}{r}\right)e^{-i\phi}\mathbf{u}_L\right],
\label{eq: E2}
\end{multline}
then we evaluate the normalized Stokes parameters $\mathbf{n} = (S_1, S_2, S_3)/S_0$, with
\begin{align}
    &S_0 = |E_R|^2 + |E_L|^2,
    &S_1 = \mathrm{Re}\{E_R E_L^*\}, \nonumber
    \\
    &S_2 = -\mathrm{Im}\{E_R E_L^*\},
    &S_3 = |E_R|^2 - |E_L|^2,
\end{align}
and the Skyrmion charge from equation (\ref{eq:nsk}).

The near field polarization texture for $\ell=1$ is shown in Fig. \ref{fig: FIG1}(a). It reveals a complete mapping of the PS. On the beam axis ($r=0$), the field exhibits pure LCP. As the radial coordinate increases, the polarization states cover all left-handed ellipticities, linear polarization and all right-handed ellipticities until the north pole, where polarization is RCP (gray scale). All orientations of the polarization ellipse or linear vibration direction are covered twice on a simple azimuthal walk around the beam center (color scale), resulting in a skyrmion of $N_{\mathrm{sk}} = 2$ in a finite radius in the transverse plane, as seen in Fig. \ref{fig: FIG1}(b). This is in contrast to standard Stokes skyrmions. The finite radius at which the PS is covered twice, i.e., the north pole is reached, is $r_{sk}=\sqrt{2|\ell|}\,\sigma$, which is $\sqrt{2}$ times the radial position of the peak of the LG bright ring. With increasing radii up to cover the entire transverse plane, the north hemisphere of the PS is covered twice down to the equator, mapped at a large radius as radial polarization. Thus, the Skyrmion number diminishes down to unity at large radii. 

This structure evolves up to the far field, simplifying into a second-order meron, as  illustrated in Fig. \ref{fig: FIG1}(c). 
On the beam axis ($r=0$), the polarization is LCP. As the radius  increases, the polarization states cover all left-handed ellipticities and twice all polarization ellipse orientations in a round about the center, ending in linear radial polarization on the equator of the PS at large $r$. As the south hemisphere of the PS is covered twice, the resulting skyrmion charge over the entire transverse plane is $N_{sk}=1$, as seen in Fig. \ref{fig: FIG1}(d). 
For vortex pump with TC $|\ell| > 1$, the polarization textures in both the near and far fields are substantially the same with the same maximum skyrmion charge of 2 and 1 over the entire transverse plane. This is because the TC difference between the RCP and LCP components in equation (\ref{eq: E2}) for SH field remains two regardless of $\ell$ (i.e. $2\ell+1$ and $2\ell-1$).

\begin{figure}
    \centering
    \includegraphics[width=1\linewidth]{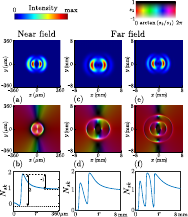}
    \caption{ (First Row) Near and far field intensity of SH field when the pump field is a LG beam ${\mathbf E}_0 \propto
    \left(r/\sigma\right)\exp\left(-r^2/2\sigma^2+i\phi\right)\mathbf{u}_x$ with TC $\ell=1$ and $\sigma =80\,\mu$m and $\lambda = 800$ nm. The inhomogeneous plasma density profile breaks the cylindrical symmetry. (a) Near field mapping of the Stokes parameters onto the transverse plane. (b) Corresponding Skyrmion charge in a disk of radius $r$. (c) and (d) the same variables after $10\, z_r$ propagation. In this example, the transverse profile of electron density is assumed to be proportional to the intensity of the pump pulse, $n_e \propto |\mathbf{E}_0|^2$. (e) and (f) the same variables after $10 \, z_r$ propagation and electron density $n_e \propto |\mathbf{E}_0|^4$.}
    \label{fig: FIG2}
\end{figure}

This analysis assumes a transversely homogeneous plasma. In real settings, the plasma generated by an intense pump pulse is inhomogeneous, as the ionization distribution follows the transverse intensity profile of the pump beam, leading to the formation of a plasma channel with a nonuniform electron density profile. Under these conditions, gradients of the electron density play a non-negligible role in the nonlinear response, and the second term in the free-electron nonlinear polarization in equation (\ref{eq: P2}) must be included. 

We assume that the transverse electron density profile can be written as $n_e(r) = n_{e,0}\, g(r/\sigma)$, where $g(r/\sigma)$ is a dimensionless function that describes the electron density distribution of the plasma channel formed by the vortex pump. Including this inhomogeneous contribution, assuming $\varepsilon_p\approx 1$, and undepleted pump, we can evaluate the second-order polarization substituting $n_e(r)$  and equation (\ref{eq: E_0}) in equation (\ref{eq: P2}), and the SH field with (\ref{eq:SH}), leading to
\begin{multline}
    \mathbf{E}_{2\omega} \propto \left(\frac{r}{\sigma}\right)^{2|\ell|}f^2e^{i2\ell\phi} \times
    \\
    \Bigg[\left(\frac{1}{f}\frac{d f}{d r}+\frac{|\ell|}{r} \right)\mathbf{u}_r + \frac{i\ell}{r}\mathbf{u}_\phi + \frac{4}{g}\frac{d g}{d r}\cos\phi\,\mathbf{u}_x\Bigg],
\end{multline}
which, expressed in the RCL and LCP basis becomes
\begin{multline}
    \mathbf{E}_{2\omega} \propto \left(\frac{r}{\sigma}\right)^{2|\ell|}f^2e^{2i\ell\phi} \times \Bigg\{
    \\
    \left[\left(\frac{1}{f}\frac{d f}{d r} + \frac{|\ell| - \ell}{r}\right)e^{i\phi} + \frac{4}{g}\frac{d g}{d r}\,\cos\phi
    \right]\mathbf{u}_R 
    \\
    + \left[\left(\frac{1}{f}\frac{d f}{d r} + \frac{|\ell| + \ell}{r}\right)e^{-i\phi} + \frac{4}{g}\frac{d g}{d r}\,\cos\phi
    \right]\mathbf{u}_L \Bigg\}.
\end{multline}
First, we assume that the electron density profile follows the pump intensity, $g(r/\sigma) \propto |\mathbf{E}_0|^2 \propto r^{2|\ell|} f^2(r/\sigma)$. As seen in Fig. \ref{fig: FIG2}, plasma inhomogeneity breaks the cylindrical symmetry, leading to a petal-like intensity pattern. Since the plasma inhomogeneity term in equation (\ref{eq: P2}) is aligned with the linear polarization direction of the pump field, this contribution introduces a preferred transverse axis that induces the deformation of the polarization texture at the near field, as in Fig. \ref{fig: FIG2}(a), the shape of which reflects the underlying plasma gradients. 

This deformation is reflected in the polarization texture, which however maintains the key features of the homogeneous case, both in the near and far fields. As seen in Fig. \ref{fig: FIG2}(a) for the near field, outside an ``exclusion'' disk around the beam center, polarization starts as LCP (white ring) in the south pole of the PS a quickly covers all left- and right-handed ellipticities up to the RCP or north pole of the PS (black ring), with twice all polarization ellipse orientations. This results in a skyrmionic polarization texture with $N_{sk}=2$ in this thin disk, as seen in Fig. \ref{fig: FIG2}(b). As in the homogeneous case, beyond the black radius, the north hemisphere of the PS is covered twice, with a deformed radial polarization distribution at large radii, resulting in $N_{sk}=-1$ from the black ring to the entire transversal plane, or $N_{sk}=1$ from the black ring. Also as in the homogeneous case, only the second-order meron structure subsists in the whole transversal plane at the far field, with $N_{sk}=1$, as seen in Figs. \ref{fig: FIG2}(c) and (d).

In Figs. \ref{fig: FIG2}(a-d), $n_e\propto |\mathbf{E}_0|^2$ is assumed as an example, but $n_e$ follows a much more complex pattern depending of the specific ionization mechanism of the gas \cite{couairon_femtosecond_2007, chin_femtosecond_2010}. Figures \ref{fig: FIG2} (e) and (f), where $n_e\propto |\mathbf{E}_0|^4$, illustrate the remarkable sensitivity of the polarization texture to the plasma density. Variations in the spatial distribution of the electron density gradient manifest as measurable changes in the shape of the polarization texture, providing an optical signature of the plasma profile and a tool for plasma diagnosis.

In conclusion, we have demonstrated that optical Stokes skyrmions can emerge spontaneously in the SH field from the nonlinear collective response of free electrons in underdense plasmas driven by optical vortices. Unlike standard Stokes skyrmions, these SH skyrmions do not require sophisticated photonic engineering, but they arise naturally from the inherent vectorial nature of SH generation in plasmas. Our results show that in a transversely homogeneous plasma, the SH field generates a complex polarization texture that fully maps the PS both in the near and far field. When plasma density gradients are taken into account, the skyrmion structure breaks its cylindrical symmetry, but maintains the full mapping of the PS. The deformation of the skyrmionic polarization texture is highly sensitive to the electron density gradients, which provides an alternative tool for plasma inhomogeneity diagnosis when conventional methods are limited.

\acknowledgments

M.G.B. and E.O. acknowledge support from project CNS2023-144058, funded by MCIN/AEI/10.13039/501100011033 and European Union "NextGenerationEU"/PRTR. M.A.P. acknowledges support from the Spanish Ministry of Science and Innovation, Gobierno de España, under Contract No. PID2021-122711NB-C21.
\bibliography{PlasmaSkyrmion}





\end{document}